\documentclass[aps,twocolumn,showpacs,amssymb,superscriptaddress]{revtex4-2}
\date{\today}
\usepackage{graphicx}
\usepackage{color}
\usepackage{amsmath}
\usepackage{array}
\usepackage{multirow}
\usepackage{longtable}
\usepackage{makecell}

\begin{document}
\title{Global systematics and theoretical interpretation of $l$-forbidden $M1$ transitions in odd-$A$ nuclei}
\author{Yu-kun Li}
\affiliation{Sino-French Institute of Nuclear Engineering and Technology, Sun Yat-sen University, Zhuhai 519082, Guangdong, China} 
\author{Guang-xin Zhang}
\email{E-mail address: zhanggx37@mail.sysu.edu.cn}
\affiliation{Sino-French Institute of Nuclear Engineering and Technology, Sun Yat-sen University, Zhuhai 519082, Guangdong, China} 
\author{Chong Qi}
\email{E-mail address: chongq@kth.se}
\affiliation{Department of Physics, KTH Royal Institute of Technology, AlbaNova University Center, SE-10691 Stockholm, Sweden}
\affiliation{Sino-French Institute of Nuclear Engineering and Technology, Sun Yat-sen University, Zhuhai 519082, Guangdong, China} 
\begin{abstract}
The $l$-forbidden magnetic dipole ($M1$) transitions, characterized by a change in orbital angular momentum ($\Delta \ell = 2$), serve as sensitive probes of higher-order effects, including configuration mixing and meson-exchange currents.
In this work, we present a comprehensive systematic study of all experimentally known $l$-forbidden $M1$ transitions, covering odd-$A$ nuclei with neutron numbers $27 \leq N \leq 126$. 
To interpret these global systematics, we apply a theoretical framework based on the relativistic Dirac wave function. This approach directly links the $l$-forbidden $M1$ transition amplitudes between pseudospin-partner orbitals to experimental single-particle magnetic moments. We perform a global comparison across isotopic chains by substituting unknown magnetic moments with rescaled Schmidt estimates.
Focusing on dominant transition groups, including $p_{3/2} \rightarrow f_{5/2}$, $s_{1/2} \rightarrow d_{3/2}$, and $d_{5/2} \rightarrow g_{7/2}$, our analysis establishes a robust linear correlation between the transition amplitudes $\sqrt{B(M1)}$ and the corresponding empirical single-particle matrix elements $M_{\mathrm{sp}}$. 
The proportionality coefficient $\rho$ serves as an empirical measure of single-particle strength fragmentation and quantifies the role of configuration mixing in driving $l$-forbidden transitions across the nuclear chart. 

\end{abstract}

\maketitle
\section{Introduction}
The nuclear magnetic dipole ($M1$) operator is fundamentally a one-body operator defined by the sum of orbital and spin angular momentum components in the form \cite{ref1}
\[
\hat{M}(M1)_{\mu}=\sqrt{\frac{3}{4\pi}}\mu_{N}\sum_{k}\left[g_{l}(k)\,\hat{\ell}_{k,\mu}+g_{s}(k)\,\hat{s}_{k,\mu}\right],
\]
where $\mu_{N}$ is the nuclear magneton, and $g_{l}$ and $g_{s}$ are the orbital and spin $g$-factors, respectively. The corresponding reduced transition probability is defined as
\[
B(M1;J_{i}\to J_{f})=\frac{1}{2J_{i}+1}\,\left|\langle J_{f}||\hat{M}(M1)||J_{i}\rangle\right|^{2}.
\]
Under standard selection rules, $M1$ transitions conserve orbital angular momentum ($\Delta \ell = 0$), yet transitions with $\Delta \ell = 2$ are frequently observed experimentally \cite{PhysRevC.110.014328,PhysRevC.93.044303,PhysRevC.102.014329,ref13,ref23}. The one-body $M1$ operator cannot connect states with $\Delta\ell\neq 0$; these transitions therefore vanish at leading order and arise only from higher-order contributions. Such transitions are commonly referred to as $l$-forbidden $M1$ transitions. Despite their significantly suppressed strengths, $l$-forbidden $M1$ transitions have been observed across a wide range of odd-$A$ nuclei and provide a sensitive probe of higher-order effects, including configuration mixing and two-body mechanisms such as meson-exchange currents and core polarization~\cite{Andrejtscheff1981-zf,ref7,ref8,ref9}.

\begin{figure*}[t] \centering \includegraphics[width=0.96\textwidth]{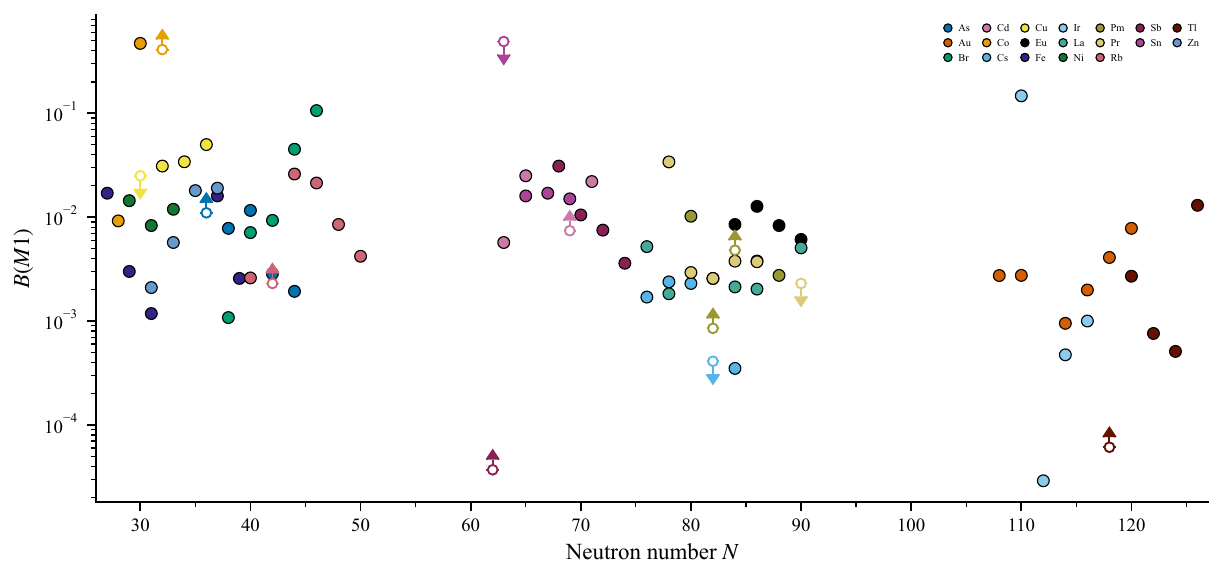} \caption{Systematics of the reduced transition probabilities $B(M1)$ for observed $l$-forbidden $M1$ transitions as a function of neutron number $N$.} \label{fig:BM1_neutron_number} \end{figure*}

The one-body $M1$ matrix element vanishes for $\Delta \ell \neq 0$ in a pure single-particle picture. The orbital term $\hat{\ell}_{k,\mu}$ connects only states with the same $\ell$, while the spin term $\hat{s}_{k,\mu}$ requires an identical spatial orbital for a nonzero radial overlap. Both therefore vanish for $\Delta\ell\neq 0$.

The observation of finite $l$-forbidden $M1$ transition strengths necessarily implies higher-order effects or many-body mechanisms. 
At the lowest order, an $l$-forbidden $M1$ transition typically involves a change in orbital angular momentum of $\Delta \ell = 2$ while the parity remains unchanged. Such transitions therefore connect single-particle orbitals with the same parity but different orbital structure, for example $s_{1/2}\leftrightarrow d_{3/2}$, $p_{3/2}\leftrightarrow f_{5/2}$, or $d_{5/2}\leftrightarrow g_{7/2}$. 
Several responsible mechanisms have been discussed in the literature and are briefly summarized below:

\paragraph*{(i) Configuration mixing}
The initial and final nuclear states are generally not of pure single-particle nature but contain admixtures of other configurations. Small components with the same orbital angular momentum $\ell$ as the partner state can therefore produce a nonzero overlap, leading to finite $M1$ transition strength. 

A minimal description of configuration mixing can be expressed as
\begin{equation}
|\Psi_{i}\rangle = a_{i} |sp_{i}\rangle + \sum_{\alpha} c_{i\alpha} |\alpha\rangle,
\end{equation}
\begin{equation}
|\Psi_{f}\rangle = a_{f} |sp_{f}\rangle + \sum_{\beta} c_{f\beta} |\beta\rangle,
\end{equation}
where $|sp\rangle$ denotes the dominant single-particle configuration and $|\alpha\rangle$, $|\beta\rangle$ represent additional configurations mixed into the wave functions. In this case, cross terms of the form $\langle \alpha | \hat{M}(M1) | \beta \rangle$ can be nonzero even when the dominant single-particle transition involves $\Delta \ell \neq 0$.
Within the large-scale shell-model framework, such effects can be treated explicitly by diagonalizing the Hamiltonian in a model space large enough to include configurations connected by the $M1$ operator across the $\Delta\ell=2$ gap. 
Here, the many-body reduced matrix element factorizes as the single-particle reduced matrix element times a spectroscopic amplitude, which approaches unity for a pure single-particle configuration. In the case of $l$-forbidden transitions, however, the single-particle matrix element is strongly suppressed, resulting in a vanishingly small transition strength.
At present, large-scale shell-model calculations can be reliably performed only for light and intermediate-mass nuclei, as well as for heavy nuclei close to magic numbers where the relevant configuration space is sufficiently limited. For nuclei far from shell closures, the rapid growth of the model space renders such calculations prohibitively demanding. As a result, a comprehensive and systematic shell-model description of $l$-forbidden $M1$ transitions over the entire nuclear chart remains beyond current computational capabilities.

\paragraph*{(ii) Two-body currents and core polarization}
Two-body contributions to the electromagnetic current, such as meson-exchange currents (MEC), generate effective $M1$ operators that do not obey the same $\Delta \ell = 0$ selection rule as the one-body operator. These two-nucleon currents can therefore directly connect configurations with $\Delta \ell \neq 0$.

In addition, virtual particle-hole excitations of the nuclear core lead to core-polarization effects, which renormalize the effective $M1$ operator through second-order processes. The resulting induced orbital and spin components may produce small but finite $l$-forbidden transition strengths. In practical shell-model calculations~\cite{ref3}, such contributions are commonly absorbed into effective $M1$ operators. Tensor terms have also been shown to play a non-negligible role in certain cases~\cite{Towner1987,Arima1987,TownerKhanna1983,BrownWildenthal1987,vonNeumannCosel2000,Andrejtscheff1981-zf,ref11,ref13}.

\paragraph*{(iii) Collective mechanism}
If the nuclear states exhibit collective or deformed character, additional $M1$ strength may arise through collective modes~\cite{ref2}.
In deformed nuclei, collective orbital currents may produce comparatively strong $M1$ transitions that are not well described by a pure spherical single-particle picture. The Nilsson model, which incorporates deformation-induced configuration mixing, provides a convenient framework for describing these effects.

Nilsson orbitals $|\Omega (N n_{z} \Lambda)\rangle$ can be expressed as linear combinations of spherical basis states:
\begin{equation}
|\Omega\rangle = \sum_{n\ell j} c_{n\ell j} |n\ell j\rangle ,
\end{equation}
where the coefficients $c_{n\ell j}$ encode admixtures of components with different orbital angular momenta.
In the deformed basis, the reduced matrix element between Nilsson states becomes
\begin{equation}
\langle \Omega_{f} || M1 || \Omega_{i} \rangle =
\sum_{n\ell j,\,n'\ell' j'}
c^{*}_{n'\ell' j'}\, c_{n\ell j}\,
\langle n'\ell' j' || M1 || n\ell j \rangle .
\end{equation}
The presence of mixed $\Delta \ell \neq 0$ components in the Nilsson wave functions thus allows otherwise forbidden transitions to acquire finite strength. This mechanism is supported by experimental observations~\cite{ref16}.

\paragraph*{(iv) Relativistic mechanism}
Relativistic corrections to the single-particle wave function can generate suppressed yet non-vanishing $M1$ matrix elements for $l$-forbidden transitions. This approach, developed in Refs.~\cite{ref4,ref5}, provides the theoretical framework for our analysis and will be detailed in the following section.

The remainder of this paper is organized as follows. Section~\ref{theo_frame} presents the theoretical framework employed in this work. Section~\ref{exp_data} presents the systematic analysis of experimental $l$-forbidden $M1$ transition data. Section~\ref{summary} provides a summary and conclusions.

\begin{figure*}[htp]
    \centering
    \includegraphics[width=0.8\textwidth]{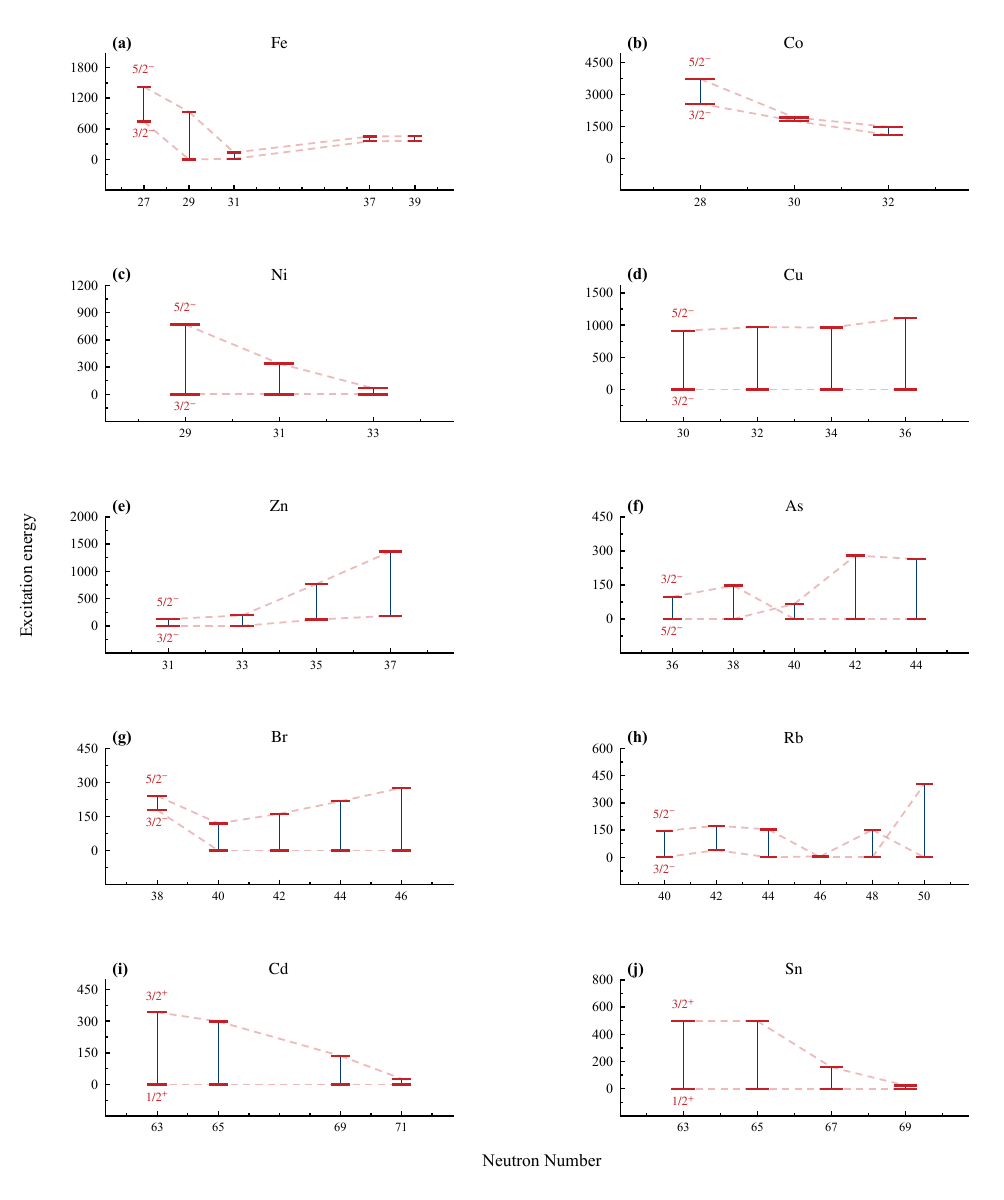}
    \caption{Summary of energy level schemes for Fe--Sn nuclei exhibiting $l$-forbidden $M1$ transitions.}
    \label{fig2}
\end{figure*}

\begin{figure*}[htp]
    \centering
    \includegraphics[width=0.8\textwidth]{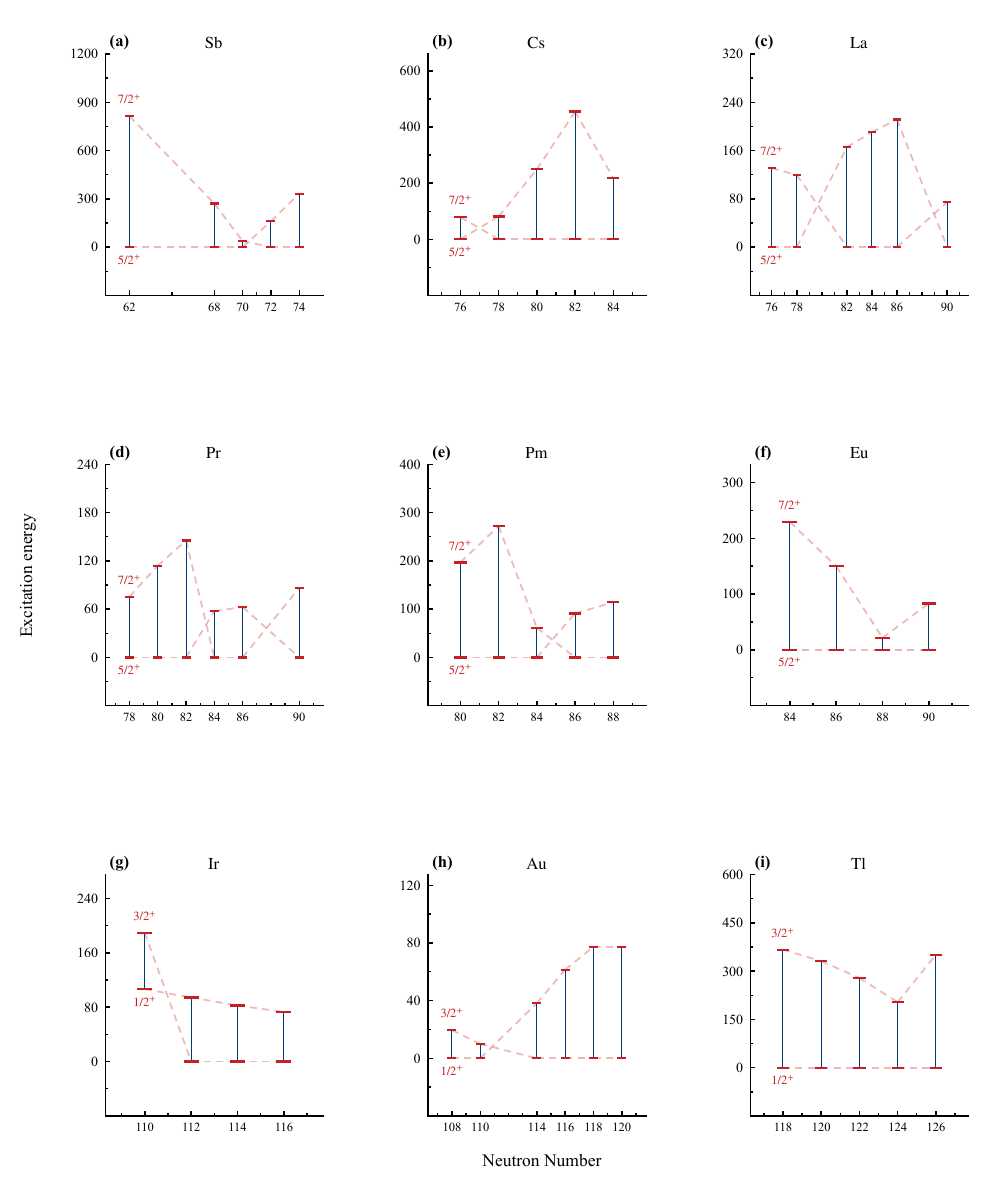}
    \caption{Same as Fig.~\ref{fig2} but for the heavier Sb--Tl nuclei.}
    \label{fig3}
\end{figure*}

\setcounter{table}{0}
\begin{table*}[htp]
\centering
\caption{Experimental $B(M1)$ values and associated single-particle magnetic moments for $l$-forbidden transitions in odd-neutron nuclei up to the Sn isotopes. The table lists the pseudospin orbital angular momentum $\tilde{l}$, the initial and final pseudospin-partner orbitals, and the corresponding experimental magnetic moments $\mu_{\exp}$ where available.}
\begin{center}
{\renewcommand{\arraystretch}{1.5}
\setlength{\tabcolsep}{10pt}
\begin{longtable}{c c c c c c}
\hline
\hline
\makecell{Nucl.} & $\tilde{l}$ & Transition & $B(M1)_{\mathrm{exp}}$ ($1.79\mu_{N}^2$) & \makecell{$\mu_{exp}$ for\\$j = l - \frac{1}{2}$ ($\mu_{N}$)} & \makecell{$\mu_{exp}$ for\\$j = l + \frac{1}{2}$ ($\mu_{N}$)} \\
\hline
$^{53}_{26}\text{Fe}_{27}$ & 2 & $f_{5/2} \rightarrow p_{3/2}$ & 0.017(5) & -0.386 & \\
$^{55}_{26}\text{Fe}_{29}$ & 2 &$f_{5/2} \rightarrow p_{3/2}$ & 0.0030(12) & & +2.7 \\
$^{57}_{26}\text{Fe}_{31}$ & 2 &$f_{5/2} \rightarrow p_{3/2}$ & 0.00118(17) & -0.1549 & +0.935 \\
$^{63}_{26}\text{Fe}_{37}$ & 2 &$f_{5/2} \rightarrow p_{3/2}$ & 0.016 & & \\
$^{65}_{26}\text{Fe}_{39}$ & 2 &$f_{5/2} \rightarrow p_{3/2}$ & 0.00257(+39-16) & & \\
$^{57}_{28}\text{Ni}_{29}$ & 2 & $f_{5/2} \rightarrow p_{3/2}$& 0.0144(18) & -0.7975 & \\
$^{59}_{28}\text{Ni}_{31}$ & 2 &$f_{5/2} \rightarrow p_{3/2}$ & 0.0083(10) & & +0.35 \\
$^{61}_{28}\text{Ni}_{33}$ & 2 & $f_{5/2} \rightarrow p_{3/2}$& 0.0119(4) & -0.75002 & +0.48 \\
$^{61}_{30}\text{Zn}_{31}$ & 2 &$f_{5/2} \rightarrow p_{3/2}$ & $\approx 0.0021$ & & \\
$^{63}_{30}\text{Zn}_{33}$ & 2 &$f_{5/2} \rightarrow p_{3/2}$ & 0.0057(+17-11) & -0.28164 & \\
$^{65}_{30}\text{Zn}_{35}$ & 2 &$f_{5/2} \rightarrow p_{3/2}$ & 0.018(+14-6) & +0.7 & +0.7690 \\
$^{67}_{30}\text{Zn}_{37}$ & 2 &$f_{5/2} \rightarrow p_{3/2}$ & 0.019(7) & +0.50 & +0.875479 \\
$^{111}_{48}\text{Cd}_{63}$ & 1 & $d_{3/2} \rightarrow s_{1/2}$& 0.0057(12) & -0.5948 & 0.012 \\
$^{113}_{48}\text{Cd}_{65}$ & 1 & $d_{3/2} \rightarrow s_{1/2}$& 0.025(8) & -0.6223 & -0.4 \\
$^{117}_{48}\text{Cd}_{69}$ & 1 &$d_{3/2} \rightarrow s_{1/2}$ & $>0.0074$ & -0.7436 & \\
$^{119}_{48}\text{Cd}_{71}$ & 1 &$d_{3/2} \rightarrow s_{1/2}$ & 0.022(4) & -0.9201 & \\
$^{113}_{50}\text{Sn}_{63}$ & 1 & $d_{3/2} \rightarrow s_{1/2}$& $<0.49$ & -0.8791 & \\
$^{115}_{50}\text{Sn}_{65}$ & 1 & $d_{3/2} \rightarrow s_{1/2}$& 0.016(3) & -0.9188 & \\
$^{117}_{50}\text{Sn}_{67}$ & 1 & $d_{3/2} \rightarrow s_{1/2}$& 0.0171(6) & -1.0010 & +0.66 \\
$^{119}_{50}\text{Sn}_{69}$ & 1 & $d_{3/2} \rightarrow s_{1/2}$& 0.015 & -1.0473 & +0.633 \\
\hline
\end{longtable}
}
\end{center}
\end{table*}

\setcounter{table}{1}
\begin{table*}[htp]
\centering
\caption{Experimental $B(M1)$ values and associated single-particle magnetic moments for $l$-forbidden transitions in odd-proton nuclei, specifically covering the $p_{3/2} \leftrightarrow f_{5/2}$ pseudospin-partner orbitals ($\tilde{l}=2$) across relevant isotopic chains.}
\begin{center}
{\renewcommand{\arraystretch}{1.5}
\setlength{\tabcolsep}{10pt}
\begin{longtable}{c c c c c c}
\hline
\hline
\makecell{Nucl.} & $\tilde{l}$ & Transition & $B(M1)_{\mathrm{exp}}$ ($1.79\mu_{N}^2$) & \makecell{$\mu_{exp}$ for\\$j = l - \frac{1}{2}$ ($\mu_{N}$)} & \makecell{$\mu_{exp}$ for\\$j = l + \frac{1}{2}$ ($\mu_{N}$)} \\
\hline
$^{55}_{27}\text{Co}_{28}$ & 2 & $f_{5/2} \rightarrow p_{3/2}$ & 0.0092(21) & & \\
$^{57}_{27}\text{Co}_{30}$ & 2 & $f_{5/2} \rightarrow p_{3/2}$ & 0.47(7) & +3.0 & \\
$^{59}_{27}\text{Co}_{32}$ & 2 & $f_{5/2} \rightarrow p_{3/2}$ & $>0.41$ & +2.54 & \\
$^{59}_{29}\text{Cu}_{30}$ & 2 & $f_{5/2} \rightarrow p_{3/2}$ & $<0.025$ & +1.8910 & \\
$^{61}_{29}\text{Cu}_{32}$ & 2 & $f_{5/2} \rightarrow p_{3/2}$ & 0.031(7) & +2.1083 & \\
$^{63}_{29}\text{Cu}_{34}$ & 2 & $f_{5/2} \rightarrow p_{3/2}$ & 0.0341(+26-22) & +2.2236 & \\
$^{65}_{29}\text{Cu}_{36}$ & 2 & $f_{5/2} \rightarrow p_{3/2}$ & 0.0499(25) & +2.3817 & +4.5 \\
$^{69}_{33}\text{As}_{36}$ & 2 & $p_{3/2} \rightarrow f_{5/2}$ & $>0.0165^*$ & & \\
$^{71}_{33}\text{As}_{38}$ & 2 & $p_{3/2} \rightarrow f_{5/2}$ & $0.0117(+51-27)^*$ & & +1.674 \\
$^{73}_{33}\text{As}_{40}$ & 2 & $f_{5/2} \rightarrow p_{3/2}$ & 0.0116(2) & & +1.63 \\
$^{75}_{33}\text{As}_{42}$ & 2 & $f_{5/2} \rightarrow p_{3/2}$ & 0.00286(11) & & +1.4398 \\
$^{77}_{33}\text{As}_{44}$ & 2 & $f_{5/2} \rightarrow p_{3/2}$ & 0.001939(4) & +1.2946 & +0.74 \\
$^{73}_{35}\text{Br}_{38}$ & 2 & $f_{5/2} \rightarrow p_{3/2}$ & 0.00108(19) & 1.97 & \\
$^{75}_{35}\text{Br}_{40}$ & 2 & $f_{5/2} \rightarrow p_{3/2}$ & 0.0071(13) & +0.76 & \\
$^{77}_{35}\text{Br}_{42}$ & 2 & $f_{5/2} \rightarrow p_{3/2}$ & 0.0093(9) & 0.92 & \\
$^{79}_{35}\text{Br}_{44}$ & 2 & $f_{5/2} \rightarrow p_{3/2}$ & 0.045(4) & +2.1064 & 2.8 \\
$^{81}_{35}\text{Br}_{46}$ & 2 & $f_{5/2} \rightarrow p_{3/2}$ & 0.106(+18-14) & +2.270562 & 1.6 \\
$^{77}_{37}\text{Rb}_{40}$ & 2 & $f_{5/2} \rightarrow p_{3/2}$ & 0.0026(12) & -0.654468 & \\
$^{79}_{37}\text{Rb}_{42}$ & 2 & $f_{5/2} \rightarrow p_{3/2}$ & >0.0023 & & \\
$^{81}_{37}\text{Rb}_{44}$ & 2 & $f_{5/2} \rightarrow p_{3/2}$ & 0.026(+27-11) & +2.0595 & \\
$^{83}_{37}\text{Rb}_{46}$ & 2 & $p_{3/2} \rightarrow f_{5/2}$ & $0.03195(25)^*$ & & +1.4249 \\
$^{85}_{37}\text{Rb}_{48}$ & 2 & $p_{3/2} \rightarrow f_{5/2}$ & $0.01275(9)^*$ & +1.35298 & \\
$^{87}_{37}\text{Rb}_{50}$ & 2 & $f_{5/2} \rightarrow p_{3/2}$ & 0.0042(+61-23) & +2.75131 & \\
\hline
\end{longtable}
}
\end{center}
\end{table*}

\setcounter{table}{2}
\begin{table*}[htp]
\centering
\caption{Experimental $B(M1)$ values and associated single-particle magnetic moments for $l$-forbidden transitions in odd-proton nuclei, specifically covering the $s_{1/2} \leftrightarrow d_{3/2}$ pseudospin-partner orbitals ($\tilde{l}=1$) observed primarily in the heavy elements Ir, Au, and Tl.}
\begin{center}
{\renewcommand{\arraystretch}{1.5}
\setlength{\tabcolsep}{10pt}
\begin{longtable}{c c c c c c}
\hline
\hline
\makecell{Nucl.} & $\tilde{l}$ & Transition & $B(M1)_{\mathrm{exp}}$ ($1.79\mu_{N}^2$) & \makecell{$\mu_{exp}$ for\\$j = l - \frac{1}{2}$ ($\mu_{N}$)} & \makecell{$\mu_{exp}$ for\\$j = l + \frac{1}{2}$ ($\mu_{N}$)} \\
\hline
$^{187}_{77}\text{Ir}_{110}$ & 1 & $d_{3/2} \rightarrow s_{1/2}$ & 0.14(7) & & +0.17 \\
$^{189}_{77}\text{Ir}_{112}$ & 1 & $s_{1/2} \rightarrow d_{3/2}$ & $0.000058^*$ & & +0.147 \\
$^{191}_{77}\text{Ir}_{114}$ & 1 & $s_{1/2} \rightarrow d_{3/2}$ & $0.000946(26)^*$ & +0.600 & +0.1507 \\
$^{193}_{77}\text{Ir}_{116}$ & 1 & $s_{1/2} \rightarrow d_{3/2}$ & $0.00200(6)^*$ & +0.519 & +0.1637 \\
$^{187}_{79}\text{Au}_{108}$ & 1 & $d_{3/2} \rightarrow s_{1/2}$ & 0.0027(4) & +0.535 & \\
$^{189}_{79}\text{Au}_{110}$ & 1 & $d_{3/2} \rightarrow s_{1/2}$ & 0.0027(5) & +0.494 & \\
$^{193}_{79}\text{Au}_{114}$ & 1 & $s_{1/2} \rightarrow d_{3/2}$ & $0.0019(34)^*$ & & 0.1396 \\
$^{195}_{79}\text{Au}_{116}$ & 1 & $s_{1/2} \rightarrow d_{3/2}$ & $0.00398(3)^*$ & & 0.1487 \\
$^{197}_{79}\text{Au}_{118}$ & 1 & $s_{1/2} \rightarrow d_{3/2}$ & $0.00818(32)^*$ & +0.420 & +0.1457 \\
$^{199}_{79}\text{Au}_{120}$ & 1 & $s_{1/2} \rightarrow d_{3/2}$ & $0.0156(3)^*$ & & +0.261 \\
$^{199}_{81}\text{Tl}_{118}$ & 1 & $d_{3/2} \rightarrow s_{1/2}$ & $>0.000061$ & +1.60 & \\
$^{201}_{81}\text{Tl}_{120}$ & 1 & $d_{3/2} \rightarrow s_{1/2}$ & 0.0027(+11-6) & +1.605 & \\
$^{203}_{81}\text{Tl}_{122}$ & 1 & $d_{3/2} \rightarrow s_{1/2}$ & 0.000759(11) & +1.622 & 0.02 \\
$^{205}_{81}\text{Tl}_{124}$ & 1 & $d_{3/2} \rightarrow s_{1/2}$ & 0.00051(10) & +1.638 & -0.8 \\
$^{207}_{81}\text{Tl}_{126}$ & 1 & $d_{3/2} \rightarrow s_{1/2}$ & 0.013(3) & +1.876 & \\
\hline
\end{longtable}
}
\end{center}
\end{table*}

\setcounter{table}{3}
\begin{table*}[htp]
\centering
\caption{Experimental $B(M1)$ values and associated single-particle magnetic moments for $l$-forbidden transitions in odd-proton nuclei, specifically covering the $d_{5/2} \leftrightarrow g_{7/2}$ pseudospin-partner orbitals ($\tilde{l}=3$).}
\begin{center}
{\renewcommand{\arraystretch}{1.5}
\setlength{\tabcolsep}{10pt}
\begin{longtable}{c c c c c c}
\hline
\hline
\makecell{Nucl.} & $\tilde{l}$ & Transition & $B(M1)_{\mathrm{exp}}$ ($1.79\mu_{N}^2$) & \makecell{$\mu_{exp}$ for\\$j = l - \frac{1}{2}$ ($\mu_{N}$)} & \makecell{$\mu_{exp}$ for\\$j = l + \frac{1}{2}$ ($\mu_{N}$)} \\
\hline
$^{113}_{51}\text{Sb}_{62}$ & 3 & $g_{7/2} \rightarrow d_{5/2}$ & $>0.000037$ & & \\
$^{119}_{51}\text{Sb}_{66}$ & 3 & $g_{7/2} \rightarrow d_{5/2}$ & 0.030(9) & +3.45 & \\
$^{121}_{51}\text{Sb}_{68}$ & 3 & $g_{7/2} \rightarrow d_{5/2}$ & 0.01047(17) & +3.36 & \\
$^{123}_{51}\text{Sb}_{70}$ & 3 & $d_{5/2} \rightarrow g_{7/2}$ & $0.01(6)^*$ & & +2.54 \\
$^{125}_{51}\text{Sb}_{72}$ & 3 & $d_{5/2} \rightarrow g_{7/2}$ & $0.0048(26)^*$ & & +2.63 \\
$^{131}_{55}\text{Cs}_{76}$ & 3 & $g_{7/2} \rightarrow d_{5/2}$ & 0.00170(5) & +3.53 & \\
$^{133}_{55}\text{Cs}_{78}$ & 3 & $d_{5/2} \rightarrow g_{7/2}$ & $0.003174(37)^*$ & +3.45 & +2.582 \\
$^{135}_{55}\text{Cs}_{80}$ & 3 & $d_{5/2} \rightarrow g_{7/2}$ & $0.003066(9)^*$ & & +2.7324 \\
$^{137}_{55}\text{Cs}_{82}$ & 3 & $d_{5/2} \rightarrow g_{7/2}$ & $\leq0.000546^*$ & & +2.8513 \\
$^{139}_{55}\text{Cs}_{84}$ & 3 & $d_{5/2} \rightarrow g_{7/2}$ & $0.000466(6)^*$ & & +2.696 \\
$^{133}_{57}\text{La}_{76}$ & 3 & $g_{7/2} \rightarrow d_{5/2}$ & 0.0052(9) & & \\
$^{135}_{57}\text{La}_{78}$ & 3 & $g_{7/2} \rightarrow d_{5/2}$ & 0.00183(6) & +3.70 & \\
$^{139}_{57}\text{La}_{82}$ & 3 & $d_{5/2} \rightarrow g_{7/2}$ & $0.003426(5)^*$ & & +2.783 \\
$^{141}_{57}\text{La}_{84}$ & 3 & $d_{5/2} \rightarrow g_{7/2}$ & $0.00284(+25-13)^*$ & & \\
$^{143}_{57}\text{La}_{86}$ & 3 & $d_{5/2} \rightarrow g_{7/2}$ & $0.002706(4)^*$ & & \\
$^{147}_{57}\text{La}_{90}$ & 3 & $g_{7/2} \rightarrow d_{5/2}$ & 0.00506 & & \\
$^{137}_{59}\text{Pr}_{78}$ & 3 & $g_{7/2} \rightarrow d_{5/2}$ & 0.034(3) & & \\
$^{139}_{59}\text{Pr}_{80}$ & 3 & $g_{7/2} \rightarrow d_{5/2}$ & 0.00293(14) & & \\
$^{141}_{59}\text{Pr}_{82}$ & 3 & $g_{7/2} \rightarrow d_{5/2}$ & 0.00257(44) & +4.2754 & +2.95 \\
$^{143}_{59}\text{Pr}_{84}$ & 3 & $d_{5/2} \rightarrow g_{7/2}$ & $0.005026(9)^*$ & +3.4 & +2.701 \\
$^{145}_{59}\text{Pr}_{86}$ & 3 & $d_{5/2} \rightarrow g_{7/2}$ & $0.00496(20)^*$ & & \\
$^{149}_{59}\text{Pr}_{90}$ & 3 & $g_{7/2} \rightarrow d_{5/2}$ & $<0.0023$ & & \\
$^{141}_{61}\text{Pm}_{80}$ & 3 & $g_{7/2} \rightarrow d_{5/2}$ & 0.0102(+15-12) & & \\
$^{143}_{61}\text{Pm}_{82}$ & 3 & $g_{7/2} \rightarrow d_{5/2}$ & >0.00085 & 3.8 & \\
$^{145}_{61}\text{Pm}_{84}$ & 3 & $g_{7/2} \rightarrow d_{5/2}$ & >0.0048 & +3.802 & \\
$^{147}_{61}\text{Pm}_{86}$ & 3 & $d_{5/2} \rightarrow g_{7/2}$ & $0.00504(12)^*$ & +3.22 & +2.58 \\
$^{149}_{61}\text{Pm}_{88}$ & 3 & $d_{5/2} \rightarrow g_{7/2}$ & $0.003667(+54-57)^*$ & +2.13 & 3.3 \\
$^{147}_{63}\text{Eu}_{84}$ & 3 & $g_{7/2} \rightarrow d_{5/2}$ & 0.0085(+10-9) & +3.736 & \\
$^{149}_{63}\text{Eu}_{86}$ & 3 & $g_{7/2} \rightarrow d_{5/2}$ & 0.0127(8) & +3.576 & \\
$^{151}_{63}\text{Eu}_{88}$ & 3 & $g_{7/2} \rightarrow d_{5/2}$ & 0.0083(4) & +3.4717 & +2.591 \\
$^{153}_{63}\text{Eu}_{90}$ & 3 & $g_{7/2} \rightarrow d_{5/2}$ & 0.00608(28) & +1.5324 & +1.81 \\
\hline
\end{longtable}
}
\end{center}
\end{table*}

\section{Theoretical framework}\label{theo_frame}
In the presence of a spherically symmetric potential, the Dirac Hamiltonian commutes with the total angular momentum $\vec{J} = \vec{L} + \vec{S}$ (associated quantum number $j$) and with the Dirac operator defined as
    \begin{equation}
    \hat{K} = \beta (\vec{\sigma} \cdot \vec{L} + 1).
    \end{equation}
For a fixed total angular momentum $j$, there are always two possible values of orbital angular momentum, $l = j \pm 1/2$.  The eigenvalues of $\hat{K}$, denoted by $\kappa$, are related to $j$ and $l$ as follows:
\begin{equation}
\kappa = 
\begin{cases} 
-(j + \frac{1}{2}) = -(l + 1) & \text{for } j = l + \frac{1}{2}, \\
+(j + \frac{1}{2}) = +l & \text{for } j = l - \frac{1}{2}.
\end{cases}
\end{equation}
Consequently, each single-particle orbital is uniquely identified by the quantum numbers $(n, \kappa)$, with $j$ and $l$ determined by $\kappa$.
The Dirac single-particle wave function is a two-component radial spinor \cite{ref20}:
\begin{equation}
\Psi_{n\kappa m}(\vec{r}) =
\frac{1}{r}
\left(
\begin{array}{c}
G_{n\kappa}(r)\,\mathcal{Y}_{\kappa m}(\hat{r}) \\
i\,F_{n\kappa}(r)\,\mathcal{Y}_{-\kappa m}(\hat{r})
\end{array}
\right).
\end{equation}
where $G_{n\kappa}(r)$ is the large (upper) radial component,
$F_{n\kappa}(r)$ is the small (lower) radial component,
and $\mathcal{Y}_{\kappa m}$ are the spinor spherical harmonics. For a given single-particle potential, the wave function can be readily evaluated with open-source codes such as those in Refs.~\cite{ref6,Amaro2025-qd}.

For a given state identified by $\kappa$, the two upper and lower components have different orbital angular momenta $l$ and $\tilde{l} = l \pm 1$, respectively.
The orbital angular momentum $\tilde{l}$ of the lower component satisfies:
\begin{equation}
\tilde{l} = 
\begin{cases} 
j - 1/2 = l - 1 & \text{if } l = j + 1/2, \\
j + 1/2 = l + 1 & \text{if } l = j - 1/2.
\end{cases}
\end{equation}
This can also be expressed using the sign of $\kappa$:
\begin{equation}
    \tilde{l} = l - \text{sgn}(\kappa).
\end{equation}

Under parity, the upper and lower components transform with opposite signs due to the $\beta$ matrix; despite carrying orbital angular momenta $l$ and $\tilde{l}$ of opposite parity, the full bispinor retains a definite parity $\pi = (-1)^l$.

We adopt the notation $j_{>}(l) = l + 1/2$ and $j_{<}(l) = l - 1/2$. In the standard non-relativistic Schr\"odinger equation framework, a single-particle state is denoted by the quantum numbers $nlj$. This corresponds to the non-relativistic limit of the Dirac single-particle state, where the small $\tilde{l}$ component vanishes. In this limit, the orbital angular momentum $l$ is a strict symmetry, and $M1$ transitions are governed by the selection rule $\Delta \ell = 0$. Consequently, transitions between states where $\Delta \ell = 2$ are strictly $l$-forbidden.

In a relativistic Dirac framework, every state possesses a small component with an orbital angular momentum $\tilde{l}$. Considering the following pairs of states where $j_{>}(l)$ transitions to $j_{<}(l+2)$, including for example $s_{1/2} \, (l=0) \rightarrow d_{3/2} \, (l=2)$, $p_{3/2} \, (l=1) \rightarrow f_{5/2} \, (l=3)$, $d_{5/2} \, (l=2) \rightarrow g_{7/2} \, (l=4)$, the upper (large) components satisfy $\Delta \ell = 2$, which leads to a vanishing $M1$ matrix element in the Schr\"odinger limit. However, the relativistic small components for these specific pairs share the same orbital angular momentum $\tilde{l} = l + 1$, thereby generating a non-zero overlap in the $M1$ transition matrix element. This coupling of $l$ and $\tilde{l}$ within the relativistic bispinor is precisely the mechanism that enables the otherwise forbidden transition. The suppressed but non-zero $M1$ strength observed in $l$-forbidden transitions is therefore a direct consequence of these small-component contributions, which allow the system to circumvent the orthogonality requirements of the non-relativistic orbital angular momentum.

In the nuclear shell model, certain single-particle states with quantum numbers $[n_{r},l,j=l+1/2]$ and $[n_{r}-1,l+2,j^{\prime}=(l+2)-1/2]$ lie close in energy. This near-degeneracy is often interpreted as a manifestation of pseudospin symmetry, which has a relativistic origin in the Dirac Hamiltonian when scalar and vector potentials nearly cancel. For example, since $d_{5/2}$ and $g_{7/2}$ both have the same $\tilde{l}=3$ in their lower components, they form a pseudospin doublet:
\[
\tilde{l}=3,\quad \tilde{s}=\frac{1}{2},\quad j=\tilde{l}\pm\frac{1}{2}.
\]
The $d_{5/2}$ and $g_{7/2}$ states are connected by the $M1$ operator even though the transition between the upper components of the pseudospin-partner wave functions is forbidden.

\subsection{$M1$ strength formulas}
Analytical expressions for single-particle $l$-forbidden $M1$ transition strengths were derived in Ref.~\cite{ref4} within a relativistic single-particle model assuming pseudospin symmetry. The fragmentation of the single-particle configuration was considered in Ref.~\cite{ref5}. In Eqs.~(1--3) of that paper, the $B(M1)$ value is corrected by the spectroscopic factors for both initial and final orbitals, and the pure single-particle magnetic moment is reconstructed by scaling the experimental moment with the inverse of the spectroscopic factor, $\mu^{\exp}/S$. While physically motivated, this form can be problematic: it relies on two spectroscopic factors that may not always be available, and the spectroscopic factor itself is not a direct observable, depending on both the choice of initial and final states and the single-particle representation. 

For the purpose of our systematic study, we propose a simple formulation where the transition amplitude is defined as the product of a one-body transition density coefficient, $\rho$, and a single-particle matrix element, $\mathcal{M}_{\mathrm{sp}}$. By assuming the transition is dominated by the same single-particle configuration that determines the ground-state magnetic properties, the $M1$ transition amplitude can be expressed directly in terms of experimental magnetic moments ($\mu^{\exp}$), as detailed below, with many-body effects and quenching absorbed implicitly through the use of measured static moments.

For odd-neutron nuclei, the $B(M1)$ value connects a doublet $j = \tilde{l} - 1/2$ and $j^{\prime} = \tilde{l} + 1/2$. Depending on the direction of the transition, two cases arise.
For the transition from the higher-lying member of the doublet ($j'$) to the lower state ($j$), the amplitude is:
\begin{equation}
\left[ B(M1; j' \to j) \right]^{1/2} = \rho \times \underbrace{ \sqrt{\frac{j+1}{2j+1}} \left( \mu_{j}^{\mathrm{eff}} - \mu_n \right) }_{\mathcal{M}_{\mathrm{sp}}}
\end{equation}
where $\rho$ is a one-body transition density coefficient representing the fragmentation of the single-particle states, $\mu_{j}^{\mathrm{eff}}$ is the measured magnetic moment of the state with total angular momentum $j$, $\mu_{n} = -1.913\,\mu_{N}$ is the magnetic moment of the free neutron, and $\sqrt{(2j'+1)/(2j+1)}$ is a statistical factor depending on which member of the doublet is higher in energy.

For the reverse transition $j \to j'$, an additional phase-space factor $\sqrt{(2j'+1)/(2j+1)}$ appears:
\begin{equation}
\left[ B(M1; j \to j') \right]^{1/2} 
= \rho \times 
\sqrt{\dfrac{2j' + 1}{2j + 1}} 
\times 
\sqrt{\dfrac{j+1}{2j+1}} 
\left( \mu_{j}^{\mathrm{eff}} - \mu_n \right)
\end{equation}

In general, the two directions are related by
\begin{equation}
\sqrt{B(M1; j \to j')} = \sqrt{\frac{2j'+1}{2j+1}}\sqrt{B(M1; j' \to j)}
\end{equation}

Equivalently, expressing the amplitude in terms of $\mu_{j'}$ gives
\begin{equation}
\left[ B(M1; j' \to j) \right]^{1/2} = \rho \times \frac{j+2}{2j+3} \sqrt{\frac{2j+1}{j+1}} \left( \mu_{j'}^{\mathrm{eff}} +\frac{j+1}{j+2} \mu_n \right)
\end{equation}

The neutron $M1$ strength is essentially proportional to the difference between the measured magnetic moment and the free neutron moment. Since $g_l = 0$ for the neutron, the transition arises entirely from the spin term.

For odd-proton nuclei, both spin and orbital contributions enter ($g_l = 1$), leading to a more involved expression. It is expressed as
\begin{widetext}
\begin{equation}
\sqrt{{B(M1;j^{\prime}\to j)}}= \rho\times \frac{(j+2)(2j+1)\mu_{j^{\prime}}^{\mathrm{eff}}-(2j+3)(j+1)\mu_{j}^{\mathrm{eff}}+4(j+1)^{2}\mu_{p}}{2(2j+3)\sqrt{{(j+1)(2j+1)}}}
\end{equation}
\end{widetext}
In this case, $\mu_{p} = 1.793 \mu_{N}$ represents the anomalous proton magnetic moment (difference between the observed value and that of a simple point-like particle with spin $1/2$).

\subsection{Schmidt moments and systematic analysis}
To investigate the correlation between $B(M1)$ values and configuration mixing, we extend our analysis using transition data compiled in the NNDC database. Where experimental magnetic moments are incomplete or unavailable, we replace them with theoretical Schmidt moments. These are defined for all initial and final state orbits as follows:
\begin{equation}
\mu^{\mathrm{Sch}} = 
\begin{cases}
(j - \frac{1}{2})g_l + \frac{1}{2}g_s & \text{for } j = l + \frac{1}{2}, \\
\frac{j}{j+1} \left[ (j + \frac{3}{2})g_l - \frac{1}{2}g_s \right] & \text{for } j = l - \frac{1}{2}.
\end{cases}
\end{equation}

\begin{equation}
\mu^{\mathrm{eff}} = 
\begin{cases}
(j - \frac{1}{2})g_l^{\mathrm{eff}} + \frac{1}{2}g_s^{\mathrm{eff}} & \text{for } j = l + \frac{1}{2}, \\
\frac{j}{j+1} \left[ (j + \frac{3}{2})g_l^{\mathrm{eff}} - \frac{1}{2}g_s^{\mathrm{eff}} \right] & \text{for } j = l - \frac{1}{2}.
\end{cases}
\end{equation}
In these expressions, $g_l$ and $g_s$ represent the orbital and spin $g$-factors, respectively, while $j$ and $l$ denote the total and orbital angular momentum quantum numbers.

In order to retain nuclei with incomplete magnetic-moment information in the systematic analysis, the effective magnetic moment entering the matrix element is defined as
\begin{equation} \mu_{\alpha}^{\mathrm{eff}} = \begin{cases} \mu_{\alpha}^{\exp}, & \text{if available},\\[2mm] q_{\alpha}\,\mu_{\alpha}^{\mathrm{Sch}}, & \text{otherwise}, \end{cases} \label{eq:effective_moment} \end{equation}
where $\alpha$ denotes the relevant orbit. The factor $q_{\alpha}$ is an empirical renormalization factor applied only to the substituted Schmidt moment. It should not be interpreted as a universal quenching parameter or a physically derived microscopic quantity.

For each nucleus, the adopted effective magnetic moments are substituted into the neutron or proton single-particle matrix element to calculate $M_{\mathrm{sp},i}$. The proportionality coefficient $\rho$ for each isotopic chain is then obtained from a least-squares fit constrained to pass through the origin,
\begin{equation}
\hat{\rho} = \frac{ \sum_i M_{\mathrm{sp},i}\sqrt{B_i(M1)} }{ \sum_i M_{\mathrm{sp},i}^{2} } .
\label{eq:rho_estimator}
\end{equation}
The fit quality is quantified by the uncentered coefficient of determination,
\begin{equation}
R_0^2 = 1- \frac{ \sum_i \left[ \sqrt{B_i(M1)}-\hat{\rho}M_{\mathrm{sp},i} \right]^2 }{ \sum_i B_i(M1) } ,
\label{eq:r2_uncentered}
\end{equation}
which is appropriate for a constrained through-origin fit testing proportionality between the transition amplitude and the single-particle matrix element. 

The empirical renormalization factors are determined independently for each isotopic chain by maximizing $R_0^2$. For odd-neutron nuclei, only the magnetic moment entering the neutron single-particle matrix element is varied when not known experimentally,
\begin{equation}
\hat{q}_{j} = \underset{q_j}{\operatorname{arg\,max}}\, R_0^2(q_j).
\label{eq:qopt_neutron}
\end{equation}
For odd-proton nuclei, the matrix element may involve both $\mu_{j'}^{\mathrm{eff}}$ and $\mu_j^{\mathrm{eff}}$, so the optimization becomes
\begin{equation}
(\hat{q}_{j'},\hat{q}_{j}) = \underset{q_{j'},q_j}{\operatorname{arg\,max}}\, R_0^2(q_{j'},q_j).
\label{eq:qopt_proton}
\end{equation} 

After the optimized effective moments are fixed, the final value of $\rho$ is extracted from Eq.~(\ref{eq:rho_estimator}) for each isotope chain. The resulting $\rho$ values are compared among nuclei belonging to the same orbital transition family. In this formulation, the central object of the systematics is the empirical proportionality between the observed amplitude $\sqrt{B(M1)}$ and the magnetic-moment-based single-particle matrix element $M_{\mathrm{sp}}$. The coefficient $\rho$ should not be identified directly with a spectroscopic factor unless it is independently constrained by reaction or transfer data.

\section{Systematics of experimental data}\label{exp_data}

Reference~\cite{ref22} presented one of the earliest systematic investigations of $l$-forbidden $M1$ transitions. Despite the limited experimental information then available, this global compilation provided the first global survey of $l$-forbidden $M1$ strengths across the nuclear chart.
Several limitations of that early analysis are, however, now apparent. The available data were sparse, and transitions from different mass regions and distinct single-particle orbitals were necessarily merged into a single discussion. Isotope- and orbital-dependent trends were therefore not resolvable, and the interpretation of the observed regularities remained qualitative.

In the six decades since, advances in experimental nuclear spectroscopy have substantially expanded the available $l$-forbidden $M1$ transition data. This enlarged body of data now permits an orbital- and isotope-resolved analysis of the underlying single-particle structure, going well beyond a global survey of transition strengths.

Our analysis proceeds in three stages. We first examine the systematics of the initial and final single-particle orbital levels along isotopic chains. We then analyze the magnetic-moment systematics of the relevant orbitals. Finally, for both odd-neutron and odd-proton nuclei, we investigate the linear correlations between $\sqrt{B(M1)}$ and the calculated single-particle matrix element $M_{\mathrm{sp}}$ within the same initial--final orbital transition scheme. By grouping nuclei with common orbital-transition types within a unified fitting framework, this analysis connects the evolution of $l$-forbidden $M1$ strengths directly to the underlying single-particle structure.

To make this comparison quantitative, we express the observed transition
amplitude in the form
\begin{equation}
    \sqrt{B_i(M1)}=\rho\,M_{\mathrm{sp},i}+\epsilon_i ,
    \label{eq:intro_rho_msp}
\end{equation}
where $i$ denotes a nucleus in a given isotopic chain, $M_{\mathrm{sp},i}$ is the
single-particle matrix element constructed from the static magnetic moments of
the relevant pseudospin-partner orbitals, $\rho$ is a phenomenological
amplitude-scaling coefficient, and $\epsilon_i$ denotes the residual deviation.

For odd-neutron nuclei, the matrix element entering Eq.~(\ref{eq:intro_rho_msp}) is the neutron expression defined earlier and the odd-proton matrix element is the corresponding proton expression. The experimental or rescaled Schmidt moments described above are used as input, and the coefficient $\rho$ quantifies the fragmentation of the ideal single-particle $M1$ amplitude.

All experimental quantities used in the present systematics were taken from the adopted values compiled in Ref.~\cite{NNDC}. The extracted data include level energies, spin-parity assignments, transition energies, branching ratios, mixing ratios (where available), lifetimes, $B(M1)$ values, and nuclear magnetic moments. This database is used here as an evaluated nuclear-data source rather than as a primary experimental publication; therefore, no $B(M1)$ values or magnetic moments were re-derived from raw spectra in the present work.

For transparency, the most relevant primary experimental studies underlying the adopted magnetic-moment information are also cited. Recent laser- and hyperfine-spectroscopy measurements provide magnetic or electromagnetic moments in several mass regions relevant to the present analysis, including $^{53}$Fe, neutron-deficient Cu isotopes, Cd isotopes and isomers, neutron-rich Zn isotopes, Sn isotopes, odd-mass Ni isotopes, Sb isotopes, Au isomers, and Tl isotopes near $N=126$~\cite{ref35,ref36,ref37,ref38,ref39,ref40,ref41,ref42,ref43,ref44}. In cases where the analyzed $M1$ strengths are associated with rotational, triaxial, or wobbling-band structures, the corresponding high-spin spectroscopic studies are cited separately, because these works primarily constrain the level schemes, transition placements, multipolarity assignments, and band interpretation rather than serving as global evaluated data tables~\cite{ref45,ref46,ref47}. This distinction avoids double counting and clarifies the provenance of each adopted quantity.

\begin{figure}[h]
    \centering
    \includegraphics[width=0.95\linewidth]{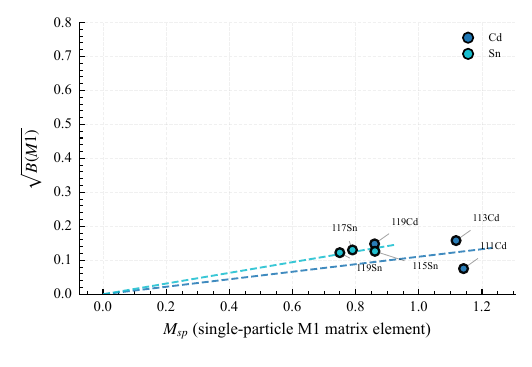}
    \caption{Correlations between $\sqrt{B(M1; 3/2^+ \rightarrow 1/2^+)}$ and $M_{\mathrm{sp}}$ across the Cd and Sn isotopic chains.}
    \label{fig4}
\end{figure}

\begin{figure}[htp]
    \centering
    \includegraphics[width=0.96\linewidth]{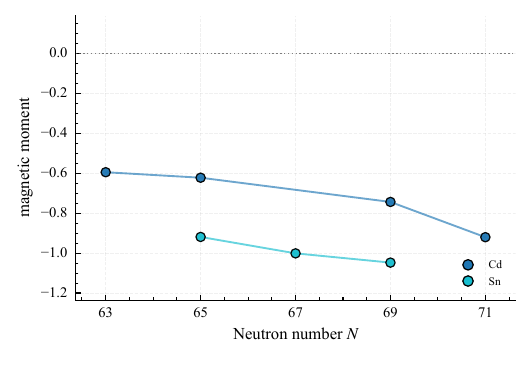}
    \caption{Evolution of the final-state magnetic moments across the Cd and Sn isotopic chains.}
    \label{fig5}
\end{figure}

\subsection{Odd-neutron nuclei, $3/2^+ \rightarrow 1/2^+$ transition}

For the $3/2^+ \rightarrow 1/2^+$ transitions in odd-neutron Cd and Sn isotopes with $N > 62$, the energy spacings between the initial and final states as shown in Figs.~\ref{fig2}(i-j) exhibit identical and relatively smooth trends. 
The correlation between $\sqrt{B(M1)}$ and $M_{\mathrm{sp}}$ is shown in Fig.~\ref{fig4}, whereas the evolution of magnetic moments is displayed in Fig.~\ref{fig5}.

Accordingly, the observed $3/2^+ \rightarrow 1/2^+$ transition can be interpreted predominantly by a single-neutron transition between $d_{3/2} $ and $ s_{1/2}$ orbits, being an $l$-forbidden $M1$ transition.
Indeed, the magnetic moments follow nearly identical trends and vary more smoothly than the energy-level separations. 

Consequently, in the plot of $\sqrt{B(M1)}$ versus $M_{\mathrm{sp}}$, with the exception of the anomalously low $B(M1)$ value for $^{111}$Cd, the data points for all other isotopes lie reasonably well along the fitting line for Cd. 
As a result, the $3/2^+\to 1/2^+$ systematics in odd-neutron Cd and Sn isotopes are thus robust overall.

\begin{figure}[htp]
    \centering
    \includegraphics[width=0.96\linewidth]{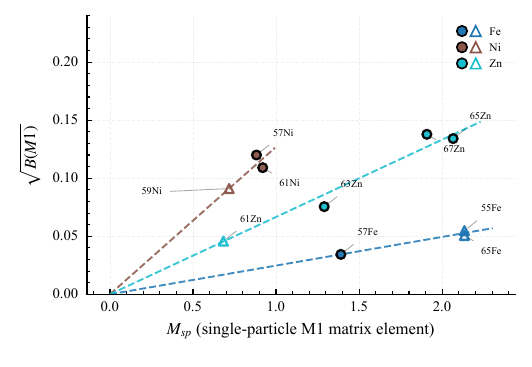}
    \caption{Correlations between $\sqrt{B(M1; 5/2^- \rightarrow 3/2^-)}$ and $M_{\mathrm{sp}}$ across the Fe, Ni, and Zn isotopic chains.}
    \label{fig6}
\end{figure}

\begin{figure}[htp]
    \centering
    \includegraphics[width=0.96\linewidth]{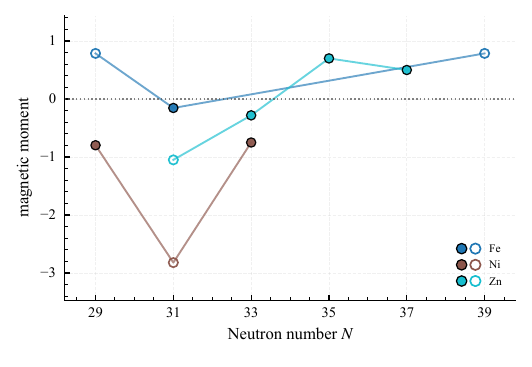}
    \caption{Evolution of the final-state magnetic moments across the Fe, Ni, and Zn isotopic chains. Filled and open markers represent the experimental and rescaled Schmidt moments, respectively.}
    \label{fig7}
\end{figure}

\subsection{Odd-neutron nuclei, $5/2^- \rightarrow 3/2^-$ transition}

For the $5/2^- \rightarrow 3/2^-$ transitions in odd-neutron Fe, Ni, and Zn isotopes, the corresponding energy-level systematics are shown in the relevant panels of Fig.~\ref{fig2}. The correlation between $\sqrt{B(M1)}$ and $M_{\mathrm{sp}}$ is displayed in Fig.~\ref{fig6}, whereas the evolution of the magnetic moments is presented in Fig.~\ref{fig7}. Among these nuclei, Fe exhibits pronounced irregular variations with neutron number, both in the absolute energies of the initial and final states and in the corresponding level spacings. By contrast, the energy-level systematics of Ni and Zn are comparatively more regular: the level spacings decrease with neutron number in Ni, while they increase in Zn.

Accordingly, the observed $5/2^- \rightarrow 3/2^-$ transitions can be interpreted predominantly as single-neutron transitions between the $\nu f_{5/2}$ and $\nu p_{3/2}$ orbitals, corresponding to an $l$-forbidden $M1$ transition. However, this interpretation is less clean than in more spherical regions. The large $\beta_2$ values in the Fe region indicate significant nuclear deformation, making it non-trivial to disentangle proton and neutron contributions to both the level energies and the orbital magnetic moments. The magnetic moments also show marked non-monotonic variations, with Fe, Ni, and Zn sharing a common minimum at $N=31$; in Zn, the magnetic moment additionally changes sign between $N=33$ and $N=35$.

In the plot of $\sqrt{B(M1)}$ versus $M_{\mathrm{sp}}$, the data points for Fe, Ni, and Zn all lie relatively close to their respective fitting lines. This indicates that the $5/2^- \rightarrow 3/2^-$ transitions retain a recognizable single-particle component despite the evident deformation effects. Nevertheless, the energy-level systematics, especially in Fe, are substantially more irregular than those in the Cd and Sn cases.

\begin{figure}[htp]
    \centering
    \includegraphics[width=0.95\linewidth]{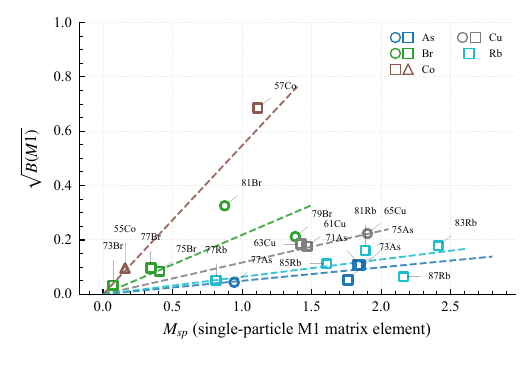}
    \caption{Correlations between $\sqrt{B(M1; 5/2^- \rightarrow 3/2^-)}$ and $M_{\mathrm{sp}}$ across the Co, Br, Cu, Rb, and As isotopic chains.}
    \label{fig8}
\end{figure}

\begin{figure}[htp]
    \centering
    \includegraphics[width=0.96\linewidth]{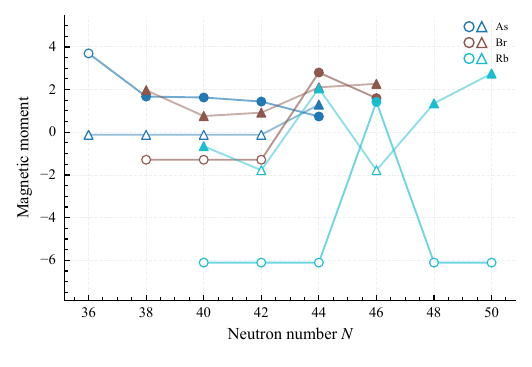}
    \caption{Evolution of the initial- and final-state magnetic moments across the As, Br, and Rb isotopic chains. Filled and open markers represent the experimental and rescaled Schmidt moments, respectively.}
    \label{fig9}
\end{figure}

\begin{figure}[htp]
    \centering
    \includegraphics[width=0.96\linewidth]{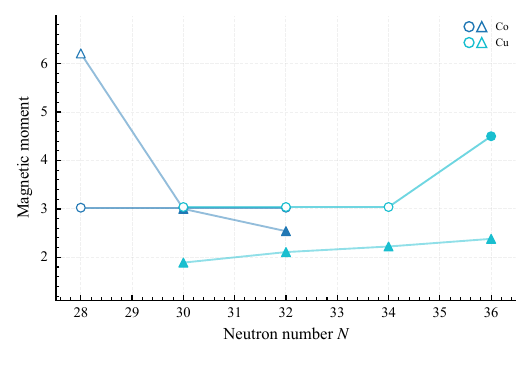}
    \caption{Same as Fig.~\ref{fig9} but for the Co and Cu isotopic chains.}
    \label{fig10}
\end{figure}

\subsection{Odd-proton nuclei, $5/2^- \rightarrow 3/2^-$ transition}

For the $5/2^- \rightarrow 3/2^-$ transitions in odd-proton Co, Cu, As, Br, and Rb isotopes, the corresponding energy-level systematics are shown in the relevant panels of Fig.~\ref{fig2}. The correlation between $\sqrt{B(M1)}$ and $M_{\mathrm{sp}}$ is displayed in Fig.~\ref{fig8}, whereas the evolution of the magnetic moments is presented in Fig.~\ref{fig9} and Fig.~\ref{fig10}. The energy-level systematics display several distinct patterns. The Cu and Br isotopic chains, except for $^{77}$Br, exhibit highly consistent and smooth trends with neutron number. For As and Rb, the level spacings between the initial and final states vary relatively smoothly, although both elements show a single level crossing between the two states. By contrast, Co does not exhibit a clear unified trend, apart from a common decrease in the absolute energies of the relevant orbitals.

Accordingly, the observed $5/2^- \rightarrow 3/2^-$ transitions can be interpreted predominantly as single-proton transitions between the $\pi f_{5/2}$ and $\pi p_{3/2}$ orbitals, corresponding to an $l$-forbidden $M1$ transition. The magnetic-moment systematics, however, are not uniform across the different isotopic chains. For Co, the magnetic moments of the final states decrease monotonically with neutron number, while for Cu, the magnetic moments of both the initial and final states increase smoothly. The cases of As, Br, and Rb are more complicated. In As, the magnetic moments of the initial and final states evolve in opposite directions, with pronounced variations at $N=42$--44, whereas in Br sizable variations appear at $N=44$--46.

In the plot of $\sqrt{B(M1)}$ versus $M_{\mathrm{sp}}$, the data points generally follow their respective fitting lines, indicating that the $5/2^- \rightarrow 3/2^-$ transitions retain a recognizable single-particle component in these odd-proton nuclei. Nevertheless, the Br isotopes show a notable anomaly: the two isotopes with experimentally confirmed orbital assignments, $^{79}$Br and $^{81}$Br, scatter more widely around the fitting line than $^{73,75,77}$Br, whose magnetic moments were obtained from the fitting procedure. As a result, the $5/2^- \rightarrow 3/2^-$ systematics in odd-proton nuclei are physically meaningful, but they are more diverse and less robust than a simple unified single-particle trend would suggest.

\begin{figure}[htp]
    \centering
    \includegraphics[width=0.95\linewidth]{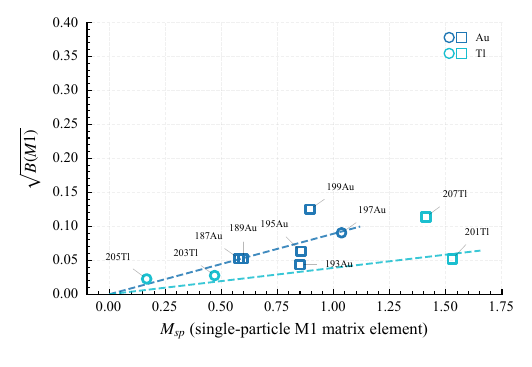}
    \caption{Correlations between $\sqrt{B(M1; 3/2^- \rightarrow 1/2^-)}$ and $M_{\mathrm{sp}}$ across the Au, Ir, and Tl isotopic chains.}
    \label{fig11}
\end{figure}

\begin{figure}[htp]
    \centering
    \includegraphics[width=0.96\linewidth]{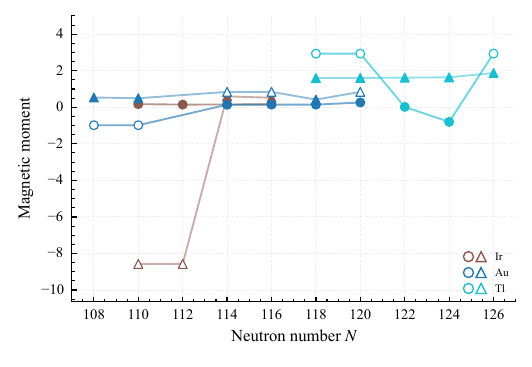}
    \caption{Same as Fig.~\ref{fig9} but for the Ir, Au, and Tl isotopic chains.}
    \label{fig12}
\end{figure}

\subsection{Odd-proton nuclei, $3/2^- \rightarrow 1/2^-$ transition}

For the $3/2^- \rightarrow 1/2^-$ transitions in odd-proton Ir, Au, and Tl isotopes, the corresponding energy-level systematics are shown in the relevant panels of Fig.~\ref{fig3}. The correlation between $\sqrt{B(M1)}$ and $M_{\mathrm{sp}}$ is displayed in Fig.~\ref{fig11}, whereas the evolution of the magnetic moments is presented in Fig.~\ref{fig12}. With the exception of a single level crossing in Ir and Au, both the absolute energies of the initial and final states and the level spacings between them vary smoothly along each isotopic chain.

Accordingly, the observed $3/2^- \rightarrow 1/2^-$ transitions can be interpreted predominantly in terms of a proton-hole transition between the $\pi 2d_{3/2}^{-1}$ and $\pi 3s_{1/2}^{-1}$ orbitals, corresponding to an $l$-forbidden $M1$ transition. The magnetic-moment systematics of Ir, Au, and Tl are particularly regular. After excluding data points flagged as preliminary, the final-state magnetic moments of Au, Ir, and Tl follow nearly identical trends. In addition, around $N=114$--116, the initial- and final-state magnetic-moment curves of Au and Ir are nearly coincident, further supporting the regularity of the systematics across these three isotopic chains.

In the plot of $\sqrt{B(M1)}$ versus $M_{\mathrm{sp}}$, the fitting lines for Tl and Ir have nearly identical slopes, indicating a highly consistent relation between the experimental transition strengths and the calculated single-particle matrix elements. The Ir data points that deviate from the fitting line correspond precisely to cases where experimental magnetic moments are unavailable, suggesting that the observed scatter is primarily associated with incomplete experimental input rather than a breakdown of the underlying systematics. As a result, the $3/2^- \rightarrow 1/2^-$ systematics in odd-proton Ir, Au, and Tl isotopes are highly robust overall. Future magnetic-moment measurements are therefore expected to further consolidate this systematics and provide a more stringent test of the proton-hole single-particle interpretation.

\begin{figure}[htp]
    \centering
    \includegraphics[width=0.95\linewidth]{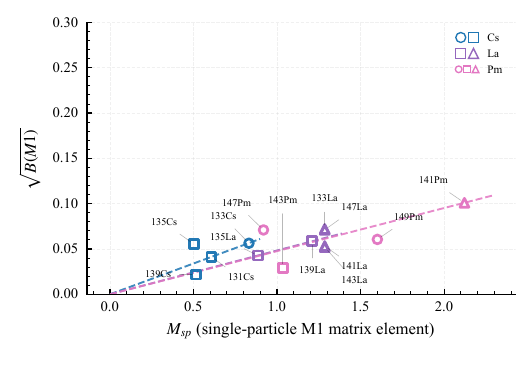}
    \caption{Correlations between $\sqrt{B(M1; 7/2^- \rightarrow 5/2^-)}$ and $M_{\mathrm{sp}}$ across the Cs, La and Pm isotopic chains.}
    \label{fig13}
\end{figure}

\begin{figure}[htp]
    \centering
    \includegraphics[width=0.95\linewidth]{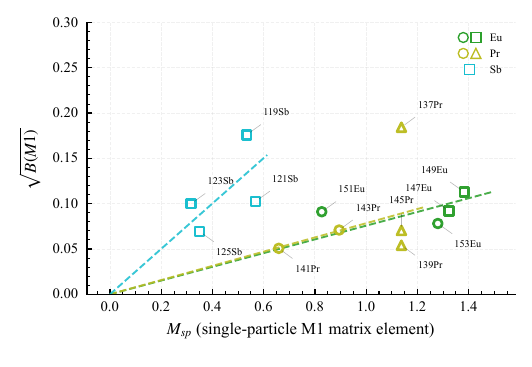}
    \caption{Correlations between $\sqrt{B(M1; 7/2^- \rightarrow 5/2^-)}$ and $M_{\mathrm{sp}}$ across the Sb, Pr and Eu isotopic chains.}
    \label{fig14}
\end{figure}

\begin{figure}[htp]
    \centering
    \includegraphics[width=0.96\linewidth]{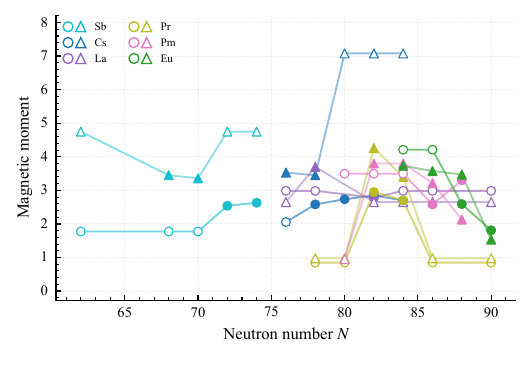}
    \caption{Same as Fig.~\ref{fig9} but for the Sb, Cs, La, Pr, Pm, and Eu isotopic chains.}
    \label{fig15}
\end{figure}

\subsection{Odd-proton nuclei, $7/2^- \rightarrow 5/2^-$ transition}

For the $7/2^- \rightarrow 5/2^-$ transitions in odd-proton Cs, La, Pr, and Pm isotopes, the corresponding energy-level systematics are shown in the relevant panels of Fig.~\ref{fig3}. The correlation between $\sqrt{B(M1)}$ and $M_{\mathrm{sp}}$ is displayed in Fig.~\ref{fig13} and Fig.~\ref{fig14}, whereas the evolution of the magnetic moments is presented in Fig.~\ref{fig15}. Unlike the more regular cases discussed above, the energy-level systematics in these nuclei are markedly irregular and vary substantially from one isotopic chain to another. Cs, La, Pr, and Pm all exhibit one or more level crossings, and the level spacings between the initial and final states fluctuate significantly with neutron number.

Accordingly, the observed $7/2^- \rightarrow 5/2^-$ transitions should be regarded as much less clean examples of $l$-forbidden $M1$ systematics than the transitions discussed in the preceding subsections. The irregular level evolution suggests that the relevant negative-parity proton configurations are strongly affected by configuration mixing and deformation effects, rather than being governed by a simple and uniform single-particle orbital transition. Consistent with this picture, both the magnetic-moment systematics and the $\sqrt{B(M1)}$--$M_{\mathrm{sp}}$ correlations show substantial chain-dependent variations. Nevertheless, the known magnetic moments associated with these orbitals remain clustered within approximately $2$--$4,\mu_N$ across the different isotopic chains.

In the plot of $\sqrt{B(M1)}$ versus $M_{\mathrm{sp}}$, the fitting behavior does not support a unified systematics comparable to that observed in the more regular odd-proton or odd-neutron cases. The data therefore indicate that the $7/2^- \rightarrow 5/2^-$ transitions in Cs, La, Pr, and Pm retain some common magnetic-moment scale, but their detailed energy and transition-strength systematics are strongly fragmented. As a result, the $7/2^- \rightarrow 5/2^-$ systematics in these odd-proton nuclei should be interpreted as limited and non-universal, rather than as a robust single-particle trend.

\section{Summary}\label{summary}
In this work, we performed a systematic investigation of $l$-forbidden $M1$ transitions across a broad range of isotopes from $N = 27$ to $126$. By analyzing the systematic trends of experimental $B(M1)$ values, we identified significant correlations between transition strengths and nuclear structure properties. To circumvent the reliance on spectroscopic factors, which are often unavailable or model-dependent, we apply a relativistic framework based on the Dirac equation, using experimental or Schmidt magnetic moments as input. 

We found a robust linear proportionality between the transition amplitudes $\sqrt{B(M1)}$ and the empirical single-particle matrix elements $M_{\mathrm{sp}}$ of the pseudospin-partner states. The derived amplitude-scaling coefficients provide an empirical measure of the fragmentation of the single-particle strengths across different isotopic chains. This systematic framework quantifies the role of configuration mixing in driving the enhancement and quenching of $l$-forbidden $M1$ transitions across the nuclear chart.


\section*{Acknowledgments}
C.Q. thanks Sun Yat-sen University for its hospitality during his visit, where this work was performed.
\section*{References}
\bibliographystyle{apsrev4-2}
\bibliography{reference}

\end{document}